\def\year{2018}\relax
\documentclass[letterpaper]{article} 
\usepackage{aaai18}  
\usepackage{times}  
\usepackage{helvet}  
\usepackage{courier}  
\usepackage{url}  
\usepackage{graphicx}  

\usepackage{times}
\usepackage{amsmath,graphicx}
\usepackage{url}
\usepackage{amsfonts}
\usepackage{amssymb}
\usepackage{verbatim}
\usepackage{subfigure}
\usepackage{epstopdf}

\usepackage{booktabs} 
\usepackage{algorithm}
\usepackage{algorithmic}
\usepackage{color}
\usepackage{amsmath}
\usepackage{amssymb}
\usepackage{subfigure}
\usepackage{graphicx}
\usepackage{epstopdf}
\newcommand{\cdcnn}{{CD-CNN}}
\newcommand{\ldcnn}{{CNN-loc}}
\newcommand{\bdcnn}{{CNN-com}}
\DeclareMathOperator*{\argmin}{argmin}

\begin{document}
%
\title{CD-CNN: A Partially Supervised Cross-Domain Deep Learning Model \\for Urban Resident Recognition}
\author{Jingyuan Wang$^1$, Xu He$^1$, Ze Wang$^1$, Junjie Wu$^{2}$, Nicholas Jing Yuan$^3$, Xing Xie$^4$, Zhang Xiong$^{5}$ \\
$^1$ School of Computer Science and Engineering, Beihang University, Beijing, China \\
$^2$ School of Economics and Management, Beihang University, Beijing, China\\
$^3$ Microsoft Corporation~~~~ $^4$ Microsoft Research \\
$^5$ Research Institute of Beihang University in Shenzhen, Shenzhen, China \\
}
\maketitle
\begin{abstract}
Driven by the wave of urbanization in recent decades, the research topic about migrant behavior analysis draws great attention from both academia and the government. Nevertheless, subject to the cost of data collection and the lack of modeling methods, most of existing studies use only questionnaire surveys with sparse samples and non-individual level statistical data to achieve coarse-grained studies of migrant behaviors. In this paper, a partially supervised cross-domain deep learning model named CD-CNN is proposed for migrant/native recognition using mobile phone signaling data as behavioral features and questionnaire survey data as incomplete labels. Specifically, CD-CNN features in decomposing the mobile data into location domain and communication domain, and adopts a joint learning framework that combines two convolutional neural networks with a feature balancing scheme. Moreover, CD-CNN employs a three-step algorithm for training, in which the co-training step is of great value to partially supervised cross-domain learning. Comparative experiments on the city Wuxi demonstrate the high predictive power of CD-CNN. Two interesting applications further highlight the ability of CD-CNN for in-depth migrant behavioral analysis.
\end{abstract}

\section{Introduction}
The continuous growth of megalopolises offer a broad range of job opportunities, which attracts a large number of migrants to seek living in the cities outside of their hometowns. According to the report of the World Bank\footnote{\url{http://data.worldbank.org/}}, in the last two decades, the urban population rate increased from 44\% to more than 53\%, implying about 5.4 hundred million people migrate from rural to cities. Driven by this urbanization wave, tremendous efforts have been dedicated to analyze and deal with social issues that caused by migration. However, limited by the channels and costs of data collection, most of existing works focus on coarse-grained migration behavior analysis on sparsely sampled questionnaire surveys and/or non-individual level statistical data~\cite{cultural_capital,rathelot2014local,marriages}.

The pretty high penetration rate of mobile phones in urban cities brings forward another way of think. Mobile phone signaling data could be used as an effective information source for demographics and urban migrants behavior analysis~\cite{mobile_phone}.
Along this line, in this paper, we propose a deep learning based cross-domain knowledge fusion framework, named Cross-Domain Convolutional Neural Network (CD-CNN), to recognize native/migrant attribute of an urban resident from mobile phone signaling data. The CD-CNN model is concerned with three core problems in mobile data-driven resident recognition:
\begin{itemize}
  \item How to extract users' behavioral features from mobile phone data. We decompose the mobile phone signaling records into two domains: the location domain and the communication domain. Convolutional neural networks are adopted to extract behavioral features from high-dimensional raw data of both domains.
  \item How to fuse knowledge from multiple domains. For heterogeneous and severely imbalanced features generated by CNNs in the location and communication domains, respectively, we introduce a carefully designed dimensionality balancing mechanism for knowledge fusion, which is crucial for the success of classification.
  \item How to handle incomplete label information of data sets. In our study, only a very small part of mobile phone users are labeled by volunteer questionnaire surveys. To deal with this, we plug a co-training scheme into the pre-training/fine-turning framework of deep learning, which solves the cross-domain learning and partially supervised learning problems simultaneously.
\end{itemize}

The proposed model offers a valuable reference to similar cross-domain data fusion scenarios, where the data have heterogeneous views, in high-dimensionality, and with incomplete label information. Comparative experimental studies on the Wuxi city with various baselines demonstrate the excellent performance of the proposed model. In particular, two interesting applications on fine-grained population census and hometown returning prediction further highlight the ability of the CD-CNN model for real-life human behavior analysis.

The remainder of this paper is organized as follows. We first describe the mobile signaling data and the volunteer survey data to be used in our study. We then propose the CD-CNN model with an emphasis on the dimensionality balancing issue. The cross-domain co-training algorithm is then designed for training CD-CNN, followed by the experiments on the Wuxi data and two real-world cases. We finally give the related work and conclude our work.

\begin{figure*}\centering
    \includegraphics[width=0.9\textwidth]{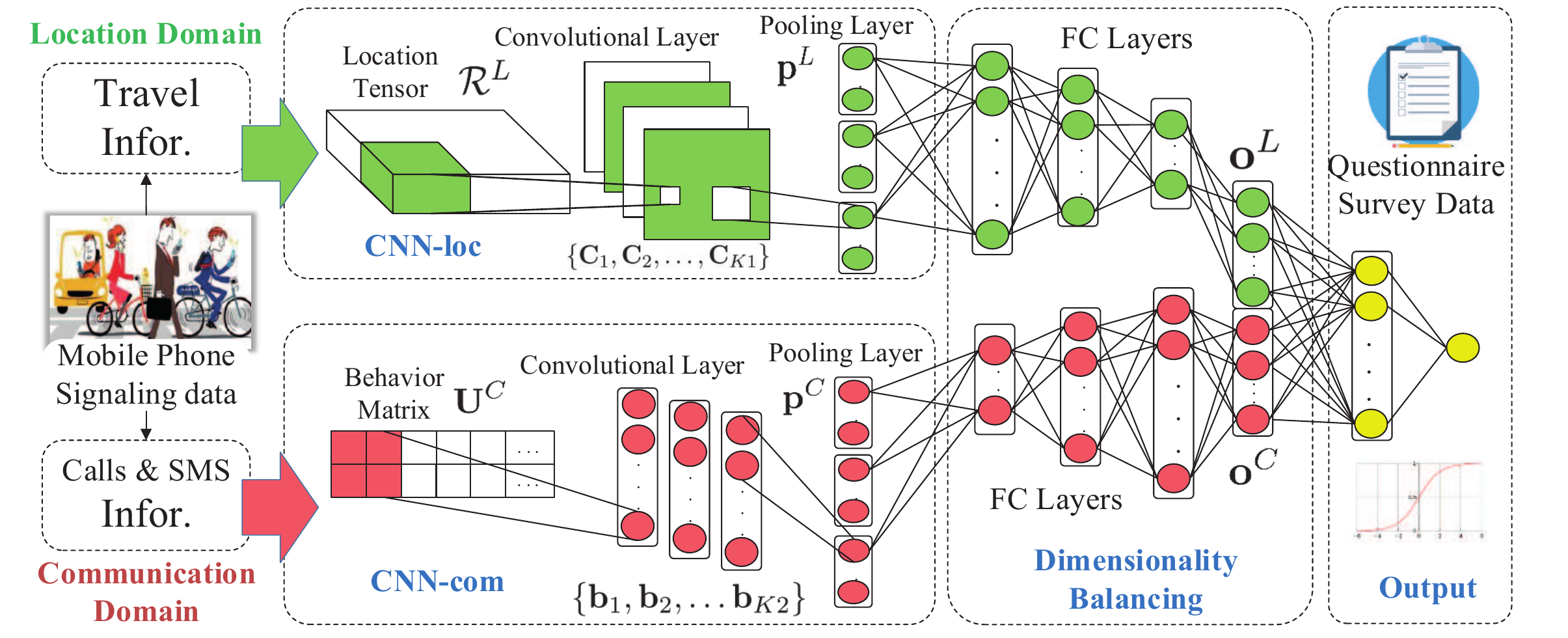}
       \caption{The framework of CD-CNN.}\label{fig:framework}
\end{figure*}

\section{Data Description}
\label{sec:data}


This work attempts to recognize an resident's identity, {\it i.e.}, a migrant or a native, through his/her activity characteristics. We define the residents who grew up out of the city they work and live in as migrants, otherwise as natives. To that end, a mobile phone signaling data set is adopted for extracting behavioral features of urban residents. Moreover, to gain some residents' ID labels for classifier training purpose, we also adopt a volunteer survey data set. Details about the two data sets are given as follows.

Mobile phone signaling data refer to the communication records between mobile phones and base stations, in which a record contains $<${\em user ID, station ID, user event code, time stamp}$>$ fields. The user ID and base station ID are unique identifications for cell phones and base stations. The user event code field records communication types of a cell phone, which include: turned on/off cell phone, started/terminated a call, sent/received a short message, connected a station or switched to another one. The time stamp records the occurrence time of a communication. From the mobile phone data set, we extract two types of behavioral information for a user, {\it i.e.}, the travel behaviors characterized by the sequence of locations the user stay in, which are approximated by the locations of base stations that the user's phone communicated with, and the communication behaviors characterized by his/her calls and SMS information through mobile phones.

The volunteer survey data set was collected from a group of randomly selected residents who voluntarily offer their mobile phone number, growth up place, and working place information in questionnaires. Using phone number of the volunteers as a foreign key to join the survey data set with the mobile phone data set, we obtain a migrant/native labeled mobile phone data set for all volunteers. Note that the questionnaires were collected anonymously, the phone numbers used to join the two data sets were hashed, and all data used in this work were authorized.

\section{The \cdcnn~Model}

In this work, we propose a partially supervised Cross-Domain Convolutional Neural Network (\cdcnn) model for accurate profiling of urban residents. Figure~\ref{fig:framework} is an illustration of \cdcnn. As shown in the figure, the model first reorganizes mobile phone data into two different domains: the location domain and the communication domain, to reflect the travel and communication behaviors, respectively. Two independent convolutional neural networks, \ldcnn~and \bdcnn, are adopted to extract resident features from the two domains. Next, a dimensionality balancing subnetwork is adopted to regularize the features extracted by \ldcnn~and \bdcnn. Finally, using an output subnetwork, the proposed model merges regularized features for classification.

\subsection{CNN for Location Domain}

The home and working places are the most important locations of a resident, using which we can infer the resident's identity. For example, a migrant worker is likely to work in an industrial park and reside in a residential area close to the industrial park. Thus, we extract home and working places information for every resident from the mobile phone data.

We divide the areas of a city into $I \times J$ square zones, and divide one day into 24 time slices. In each time slice, if there is a signaling record for a resident communicated with a based station in zone $(i,j)$, we count one for the resident in zone $(i,j)$ during that time slice. In the next step, we divide the time slices into two periods, {\it i.e.}, the working period from 7:00 to 19:00, and the home period from 19:00 to 7:00 of the next day. We define two location matrices $\mathbf{R}^h \in \mathbb{R}^{I\times J}$ and $\mathbf{R}^w \in \mathbb{R}^{I\times J}$ for a resident, where the element $r_{ij}^w$ of $\mathbf{R}^w$ is the hours the resident appeared in zone $(i,j)$ during the working period normalized by the total hours of the working period, and $r_{ij}^h$ of $\mathbf{R}^h$ is defined analogously.

The location features in $\mathbf{R}^h$ and $\mathbf{R}^w$ have two characteristics. First, the location intensity distributed in the two matrices have significant locality characteristics; that is, in most cases, a resident lives and works in a very local area of a city. It is unlikely for a resident who travels uniformly all over a city every day. Therefore, it is unnecessary to treat all elements in $\mathbf{R}^h$ and $\mathbf{R}^w$ equally. Second, the dimensionality of the two matrices is very high. Take the Wuxi city, which is an example city in our experiments below, for example, the city areas are divided into 10,120 zones.
Since the geographical features for residential and working places in different zones are very diversified, it is very hard and inefficient to reduce dimensionality of $\mathbf{R}^h$ and $\mathbf{R}^w$ using a handcrafted method. Therefore, in consideration of the characteristics of the two matrices, we adopt convolutional neural networks (CNN)~\cite{cnn}, which can extract locality features from matrix type high dimensional data, to process $\mathbf{R}^h$ and $\mathbf{R}^w$.

The proposed CNN structure consists of a convolutional layer and a pooling layer. The convolutional layer connects $\mathbf{R}^h$ and $\mathbf{R}^w$ with several trainable filters, each being a $M \times N \times 2$ weight tensor. We denote the $k$-th filter of the location domain as $\mathcal{A}_k^{L}$. The convolution layer uses $\mathcal{A}_k^{L}$ to zigzag scan a location tensor composed by $\mathcal{R}^{L} = [\mathbf{R}^h; \mathbf{R}^w]$ for the calculation of  a convolution neuron matrix $\mathbf{C}_k$, of which the $(p,q)$ convolution neuron generated by the filter $k$ is calculated by
\begin{eqnarray}\label{} 
   c_{pq}   &=& \sigma \left( u + \sum_{m=1}^M \sum_{n=1}^N  a_{nm}^h r_{p+n,q+m}^h \right. \nonumber\\
    &&\left.~~~~+\sum_{m=1}^M \sum_{n=1}^N a_{mn}^w r_{p+m,q+n}^w \right),
\end{eqnarray}
where $u$ is a trainable bias for the filter $k$, $a_{mn}^{h}$ and $a_{mn}^{w}$ are the elements of the fiber $m,n$ of $\mathcal{A}_k^{L}$, {\it i.e.}, $\left<a_{mn}^{h}, a_{mn}^{w}\right> = \mathbf{a}_{mn:}^{L}$, and $\sigma(\cdot)$ is an activation function.

A pooling layer is then adopted to reduce the dimensionality of $\mathbf{C}_k$ through an average down sampling. The pooling layer divides $\mathbf{C}_k$ into $D \times D$ disjoint regions, and uses the averages of each region to represent the convolution neurons in the region. In this way, the dimensionality of $\mathbf{C}_k$ processed by the pooling layer is reduced to $\frac{1}{D\times D}\times 100\%$ of its original size. The output of the pooling layer, denoted as $\mathbf{p}^L$, is a feature vector down sampled from the convolution neuron matrices $\{\mathbf{C}_1, \mathbf{C}_2, \ldots, \mathbf{C}_{K1}\}$.

\subsection{CNN for Communication Domain}

In the modern society, the mobile phone is becoming the most important channel for a person to contact with his/her social relations. So the mobile phone communication behaviors could effectively reflect the ways of living of a resident, which give important clues for  identity recognition.

From the mobile phone signaling data, we extract two types of operation behavior information, {\it i.e.}, calls and short messages. For a mobile phone user in the time slice $t$, we calculate $e_t$ and $s_t$ as the number of the calls and short messages normalized by total call and SMS volume of a user, respectively. A communication matrix of a resident is expressed as
\begin{equation}\label{}
\mathbf{U}^{C} = \left[
\begin{matrix}
    e_1, &e_2, &\ldots, &e_t, &\ldots, &e_{24} \\
    s_1, &s_2, &\ldots, &s_t, &\ldots, &s_{24}
\end{matrix}\right].
\end{equation}

Note that the elements in the matrix $\mathbf{U}^C$ also have the locality characteristics. Moreover, the quantities of calls and short messages during adjacent time slices have some correlations. Therefore, similar to the case of location domain, the communication matrix $\mathbf{U}^C$ is also  modeled by a convolutional neural network. The convolutional layer in the communication domain connects the communication matrix with several $2 \times m$ trainable filters. The $k$-th trainable weight filter is defined as
\begin{equation}\label{}
\mathbf{A}_k^{C} = \left[
\begin{matrix}
    a^e_1, &a^e_2, &\ldots, &a^e_H \\
    a^s_1, &a^s_2, &\ldots, &a^s_H
\end{matrix}\right].
\end{equation}
The convolution layer uses $\mathbf{A}_k^{C}$  to scan $\mathbf{U}^C$ for the generation of a convolution neuron vector $\mathbf{b}_k$. The $n$-th element of the convolution neuron vector is calculated as
\begin{equation}\label{equ:full_connect}
\begin{aligned}
    b_{n}   = \sigma \left( v + \sum_{h=1}^H a_{h}^e e_{n+h} + \sum_{h=1}^H a_{h}^s s_{n+h} \right),
\end{aligned}
\end{equation}
where $v$ is a trainable bias for the filter $k$.

Similarly as in the location domain, a pooling layer is adopted to reduce the sizes of the convolution neuron vectors generated by the convolutional layer. The output of the pooling layer is a feature vector $\mathbf{p}^C$ that are down sampled from the convolution neuron vectors $\{\mathbf{b}_1, \mathbf{b}_2, \ldots \mathbf{b}_{K2}\}$.

\subsection{Dimensionality Balancing}

Through the convolutional neural networks \ldcnn\;and \bdcnn\;mentioned above, behavioral features of residents could be extracted from the two different domains. The next step is to fuse the features of the two domains together for classification, in which the dimensionality imbalance problem emerges as an obstacle. That is, the dimensionality of features extracted from the two domains are not in the same order of magnitudes. The feature dimensionality of the location domain is about $\frac{10120}{D\times D}$ for the Wuxi city, which is much larger than that of the communication domain, which is about 24 divided by the size of a pooling window. If we directly merge the outputs $\mathbf{p}^L$ and $\mathbf{p}^C$ of \ldcnn\;and \bdcnn\; as a feature vector for resident classification, the low-dimensional communication features would be submerged by the high-dimensional location features, especially in the error back propagation algorithms where prediction errors are divided by every feature.

To meet the above challenge, we incorporate a dimensionality balancing subnetwork into the \cdcnn\;model. This network uses Fully Connected (FC) layers to adjust the number of output neurons for the two domains. As shown in Fig.~\ref{fig:framework}, the FC network connected with the high-dimensional location features $\mathbf{p}^L$ halves its neuron numbers layer by layer. On the contrary, the FC network connected with the communication features $\mathbf{p}^C$ doubles its neurons layer by layer. When the neuron numbers of the two FC networks are halved/doubled to the same order of magnitudes, we set the numbers of neurons as the same. In this way, the feature dimensions of the two domains are balanced as the same number. The feature vectors generated by the dimensionality balancing network are denoted as $\mathbf{o}^L$ and $\mathbf{o}^C$, respectively,  for the location and communication domains.


\subsection{Cross-domain Features Fusion}

In the output subnetwork, we adopt a fully connected layer to fuse cross-domain features. Using $o^L_p$ and $o^C_q$ to denote $p$-th and $q$-th elements of $\mathbf{o}^L$ and $\mathbf{o}^C$, respectively, the $k$-th neuron of the fully connected layer is given as
\begin{equation}\label{equ:full_connect}
    f_k = \sigma \left( z_k +\sum_{p} a_{pk}^L o^L_p + \sum_{q} a_{qk}^C o^C_q \right),
\end{equation}
where $z_k$, $a_{pk}^L$ and $a_{pk}^C$ are the trainable bias and parameters for the features to be fused. Finally, a logistic regression classifier is trained on the fused features $\mathbf{f} = \{f_1, f_2, \ldots\}$ for resident recognition.

\section{Cross-domain Co-training for CD-CNN}

The CD-CNN model mentioned above is fully supervised with the assumption that all the training set labels are available. However, this is not true in many real-life cases, where to label a sample is very costly. For example, in our Wuxi case, we have only 30 thousands residents whose native/migrant labels are available from the volunteer questionnaire data. In contrast, it is very easy for a mobile operator to collect mobile phone data of millions of users. Therefore, if we only use the labeled samples to train the model, it will lead to a huge waste of information for the unlabeled data.


In order to exploit the information in both labeled and unlabeled data, we propose a partially supervised network training algorithm based on the co-training scheme, named as Cross-domain Network Co-training (CNC). The CNC algorithm contains three training steps: domain separated pre-training, domain crossed co-training, and supervised fine-tuning.

For convenience, we define a prediction network model ${y}^L = LN(\theta^L, \mathcal{R}^L)$, which consists of the location domain parts of \cdcnn, {\it i.e.}, \ldcnn, feature balancing FC layers for the location domain and a logistic regression, where ${y}^L$ is a prediction output, $\theta^L$ denotes the trainable parameters of the network. Similarly, the network model consists of the communication domain parts of \cdcnn~is denoted as ${y}^C = CN(\theta^C, \mathbf{U}^C)$. In the domain separated pre-training, CNC uses labeled samples to respectively train the $LN$ model and the $CN$ model to obtain the optimized parameters denoted as $\theta^L(0)$ and $\theta^C(0)$.

In the domain crossed co-training step, the CNC algorithm uses unlabeled samples to collate the $LN$ and $CN$ models each other in an iterative way. In the $t$-th round, the algorithm selects a batch of unlabeled samples with higher prediction confidences when used as inputs of $LN$ rather than $CN$. The selected samples are then used to update the parameters $\theta^C(t-1)$ of $CN$ in the $t$-th round as
\begin{equation}\label{equ:bn_update}
    \theta^C(t) = \argmin_{\theta^C} \sum_k \left({y}^L_k(t-1) - {CN}(\theta^C, \mathbf{U}^C_k)\right)^2 + \|\theta^C\|^2_2,
\end{equation}
where ${y}^L_k(t-1) = LN(\theta^L(t-1), \mathcal{R}^L_k)$ is the label of sample $k$ predicted by the $LN$ model using a parameter generated in the ($t$-1)-th round.

Next, the algorithm repeats the same process to update the $LN$ network. A batch of unlabeled samples with a higher prediction confidences in $CN$ are selected. The parameters $\theta^L(t-1)$ of $LN$ is updated as
\begin{equation}\label{equ:ln_update}
    \theta^L(t) = \argmin_{\theta^L} \sum_{k} \left({y}^C_k(t) - LN(\theta^L, \mathcal{A}^L_k)\right)^2 + \|\theta^L\|_2^2,
\end{equation}
where ${y}^C_k(t) = CN(\theta^C(t), \mathbf{U}^C_k)$ is the label predicted by the $LN$ model using a parameter generated by \eqref{equ:bn_update}. Using the equations in \eqref{equ:bn_update} and \eqref{equ:ln_update}, the domain crossed co-training step iteratively updates $\theta^C$ and $\theta^L$ until convergence or all unlabeled data are selected.

In the fine-tuning step, the CNC algorithm once again uses labeled samples to train the model. We denote the parameter set of the \cdcnn~as
\begin{equation}\label{equ:theta}
\theta = \left\{ \theta^L, \theta^C, \theta^O \right\},
\end{equation}
where $\theta^O$ denotes the parameters of the output layers. The algorithm uses $\theta^L$ and $\theta^C$ generated by the co-training step to set the initial values of $\theta$ in \eqref{equ:theta}. Then the parameters in $\theta$ are fine tuned as
\begin{equation}\label{}
\theta = \argmin_{\theta}  \sum_{k} \left(\tilde{y}_k - \hat{y}_k\left(\mathbf{U}^C_k,  \mathcal{R}^L_k, \theta\right)\right)^2 + \|\theta\|^2_2,
\end{equation}
where $\tilde{y}_k$ is the real label of the sample $k$, and $\hat{y}_k$ is the predicted label of sample $k$ given by \cdcnn\;using $\theta$.

The three-step training of the CNC algorithm has two strengths: $i$) the knowledge from both the location and communication domains are transferred to each other, and $ii$) the knowledge hidden in unlabeled samples are exploited by the model.

\begin{figure*}\centering
	 \subfigure[Precision]{\includegraphics[width=0.3\textwidth]{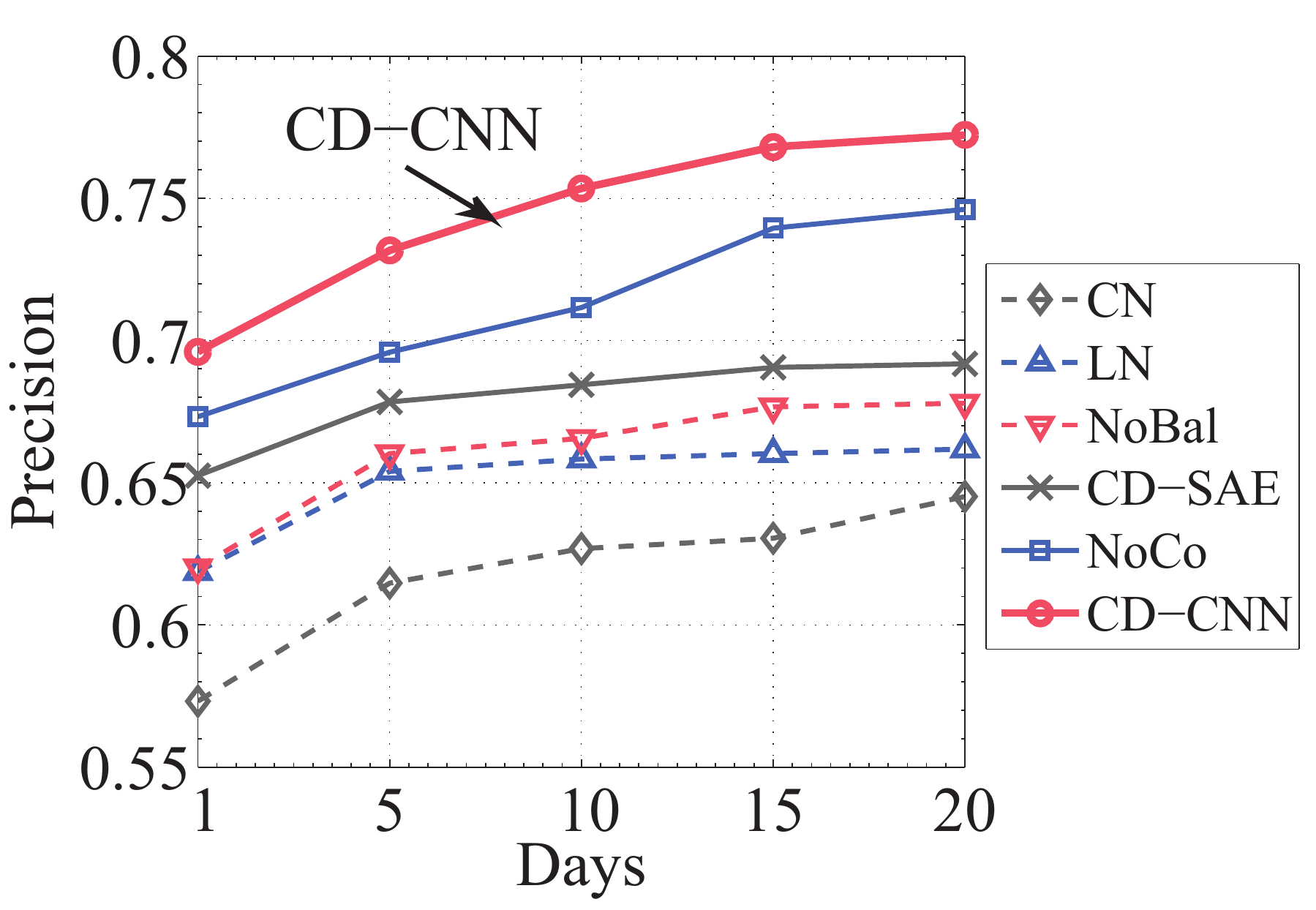}\label{fig:dayNumForPrecision}}
	 \subfigure[Recall]{\includegraphics[width=0.3\textwidth]{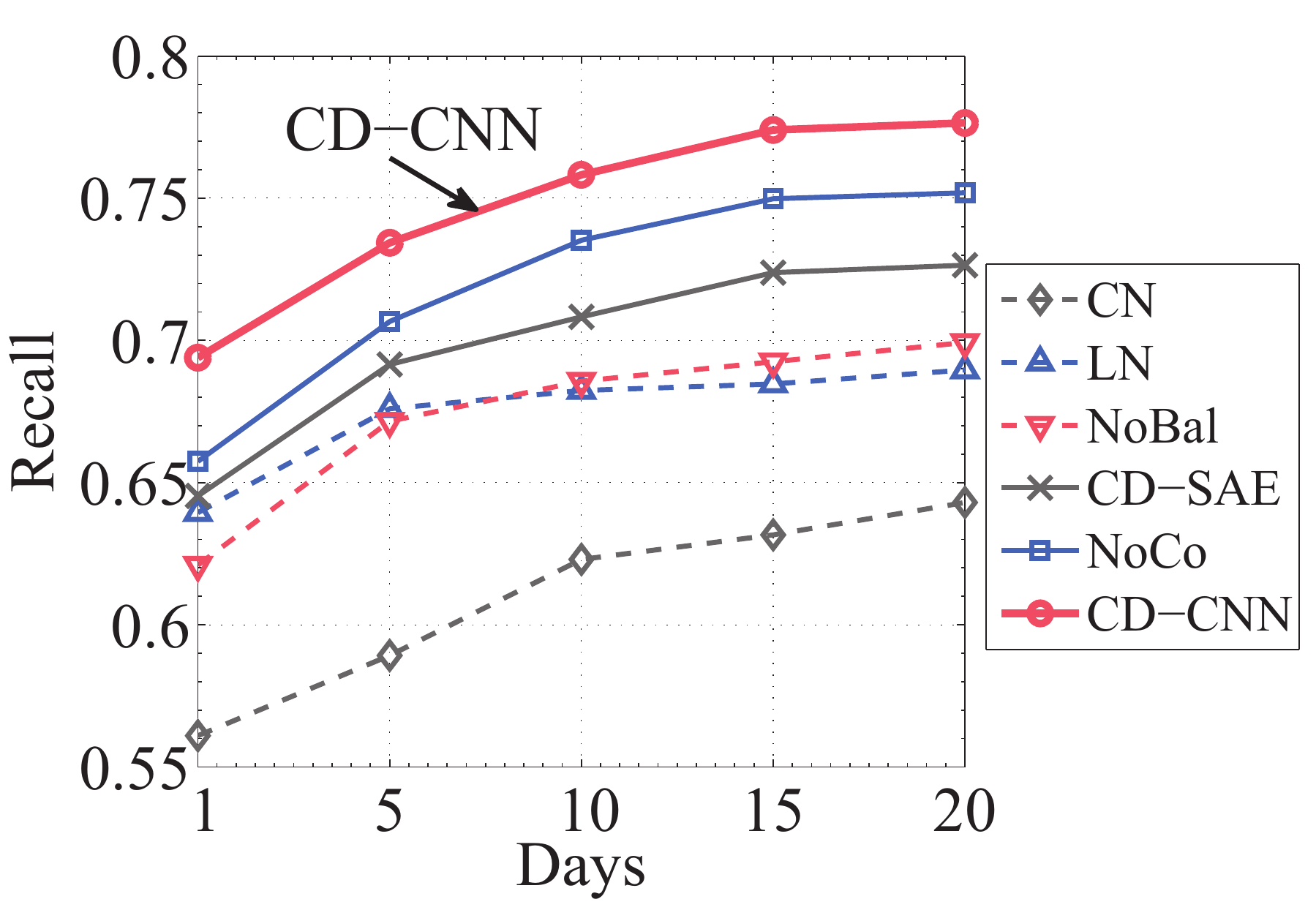}\label{fig:dayNumForRecall}}
	\subfigure[F1 Score]{\includegraphics[width=0.32\textwidth]{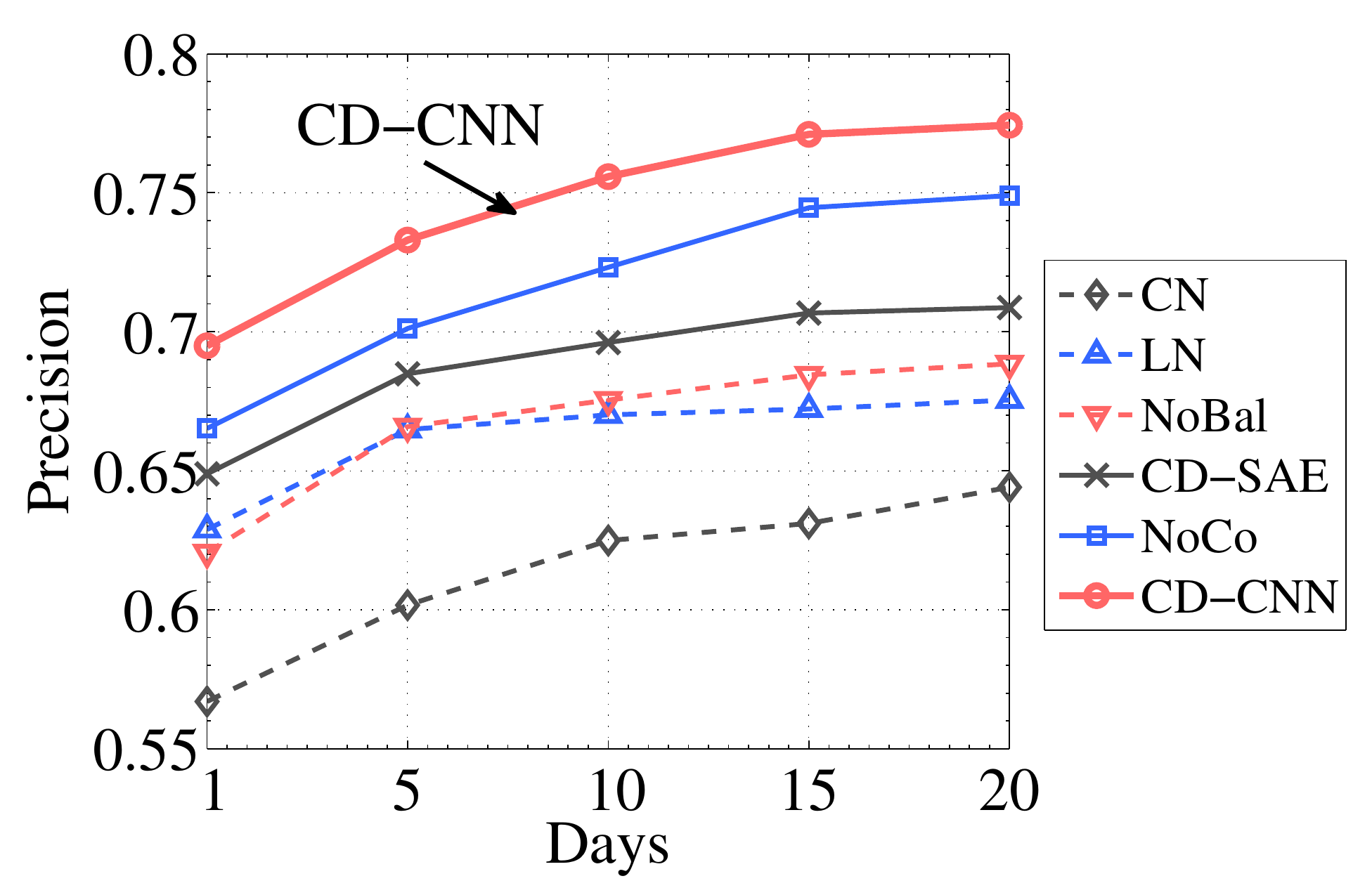}\label{fig:dayNumForFScore}}
	\caption{Classification with varying data collecting days.}
	\label{fig:differentDays}
\end{figure*}

\begin{figure*}\centering \subfigure[Precision]{\includegraphics[width=0.3\textwidth]{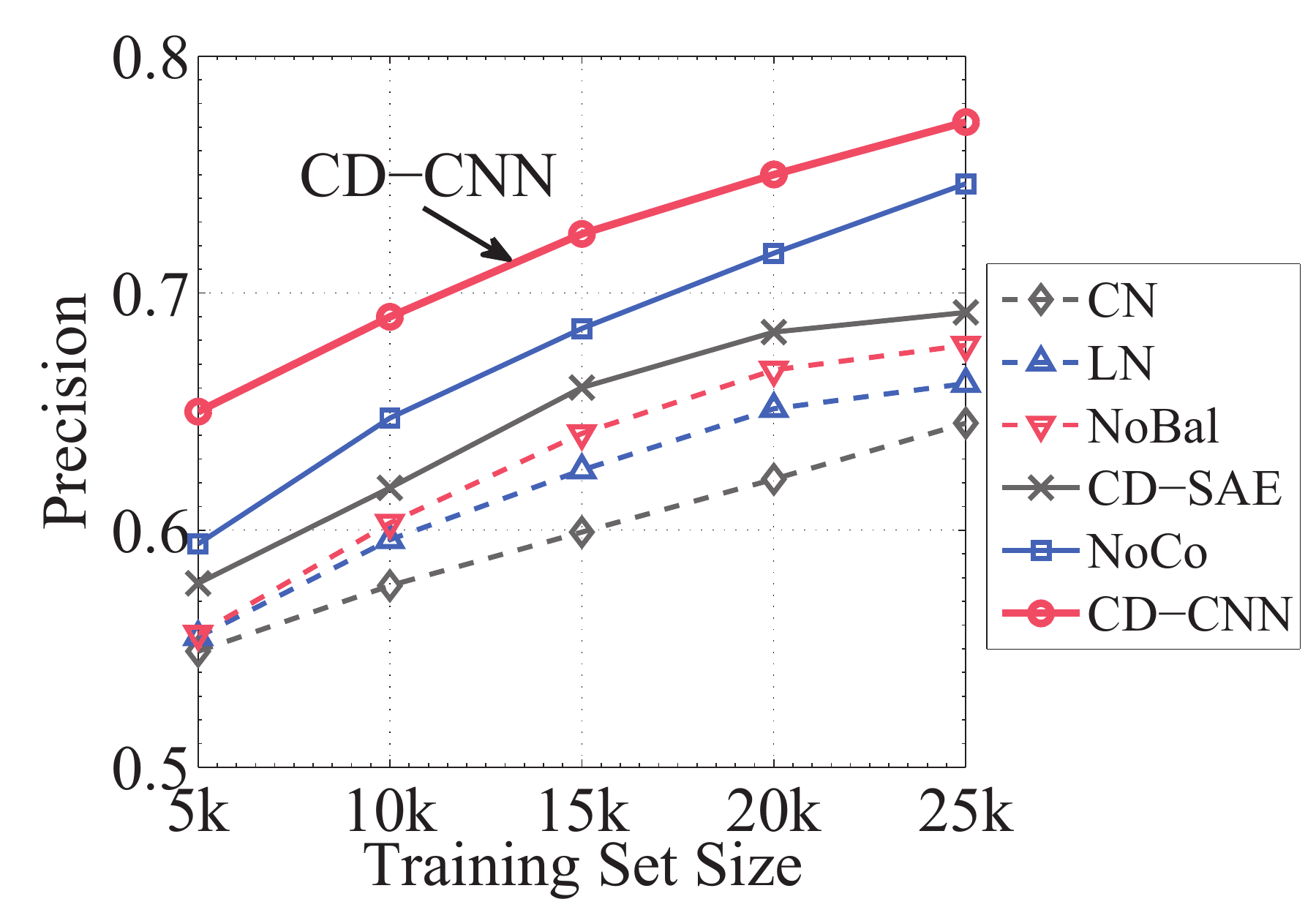}\label{fig:samplePrecision}} \subfigure[Recall]{\includegraphics[width=0.3\textwidth]{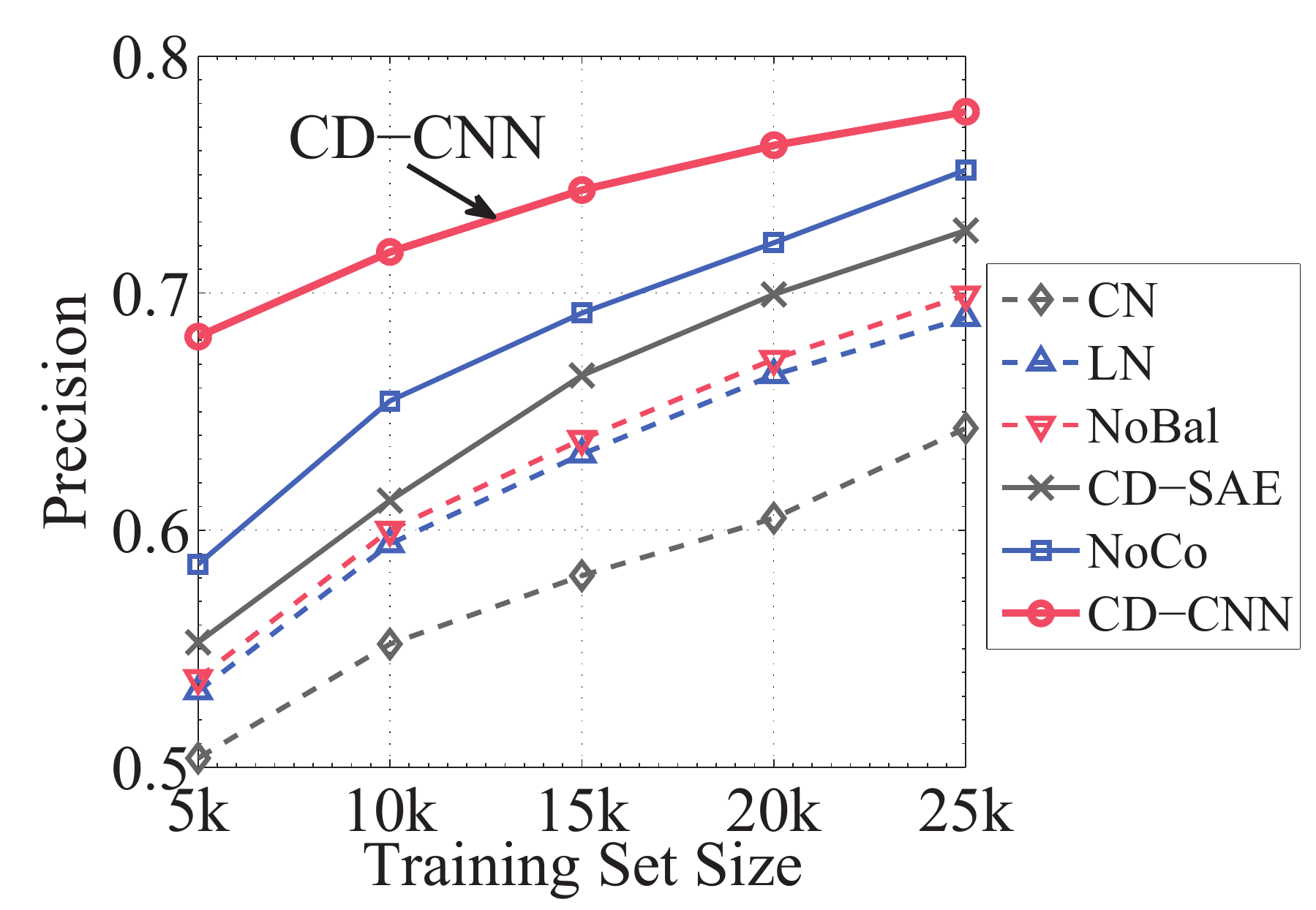}\label{fig:sampleRecall}} \subfigure[F1 Score]{\includegraphics[width=0.32\textwidth]{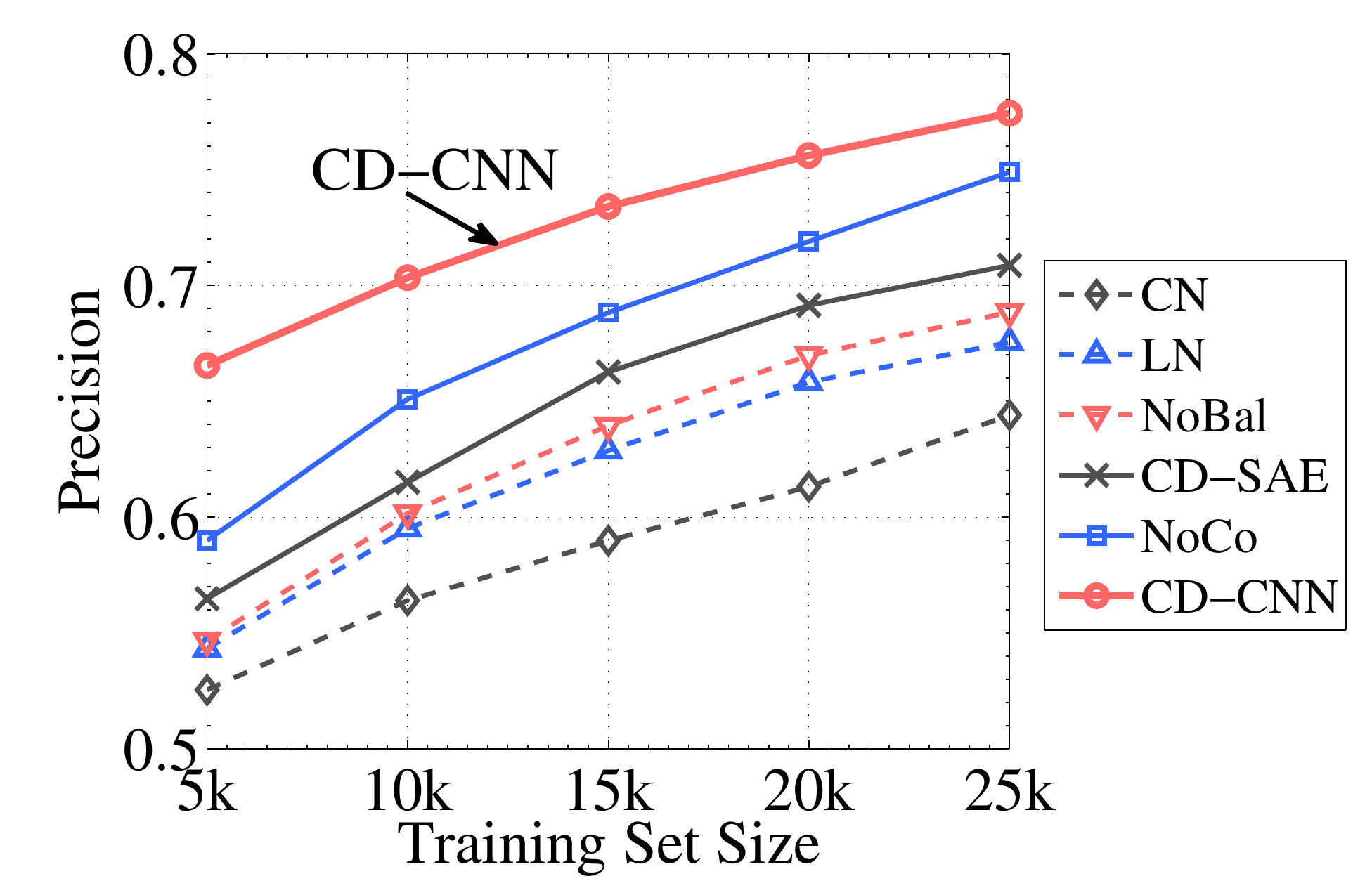}\label{fig:sampleScore}}
	\caption{Classification with varying sizes of labeled samples.}
	\label{fig:differentSet}
	\vspace{-0.3cm}
\end{figure*}

\section{Experiments}

\subsection{Experimental Setup}

The data set used in our experiments were collected from Wuxi\footnote{\url{https://en.wikipedia.org/wiki/Wuxi}}, a medium-sized city in eastern China. The population of Wuxi is about 6.5 millions, and the city area is 4,787 $km^2$. Wuxi has well-developed manufacturing industry and various large industrial parks, which attract vast migrants working and living in the city.


The mobile phone signaling data contain records about five million mobile phone users from October 2013 to March 2014. The volunteer survey data contains the information of 30 thousands volunteers, in which 50\% are natives and 50\% are migrants. Therefore, after combination the data set contains 5 million resident samples with only 30 thousands labeled. Because the city shape is an irregular polygon, we divide the circumscribing square of Wuxi into $88 \times 115 = 10,120$ square zones, each with a size of $1km \times 1km$. Altogether 3,475 zones in the circumscribing square contain base stations. We set the elements of the input tensor $\mathcal{A}^L$ that corresponds to a zone without base stations as zeros.


\subsection{Classification Performance}

We evaluate the performance of the proposed \cdcnn\; model by comparing it with several benchmarks including:
\begin{itemize}
  \item The $LN$ model, which uses the network model ${y}^L = LN(\theta^L, \mathcal{R}^L)$ to classify residents. Only the information in the location domain is exploited by this benchmark.
  \item The $CN$ model, which uses the network model ${y}^C = BN(\theta^C, \mathbf{U}^C)$ to classify residents. Only the information in the communication domain is exploited by this benchmark.
  \item No Balancing Network (NoBal), which removes the dimensionality balancing subnetwork from \cdcnn. NoBal is used to evaluate the significance of the balancing subnetwork.
  \item No Co-training (NoCo), which only uses the labeled samples to train \cdcnn. Because unlabeled samples are out of use, the benchmark does not adopt the co-training step in parameters training. NoCo is used to evaluate the significance of the co-training step.
  \item CD-SAE, which replaces the CNN structure of \cdcnn\;by Stacked Auto-Encoders (SAE)~\cite{sae}. CD-SAE is used to evaluate the effectiveness of CNN in the proposed model.
\end{itemize}

The experiments use 25 thousands labeled samples as a training set and the remaining data as the validation set. 10\% unlabeled samples, {\it i.e.}, 0.5 millions, are used in the co-training step. The precision, recall and F1-score are used as evaluation measures. Figure~\ref{fig:differentDays} plots experimental results of the proposed model and the benchmarks with incremental data-collecting days (working days only) for robustness check. 
It can be seen from Fig.~\ref{fig:differentDays} that: $i$) The performance of $LN$ is much better than $CN$, which implies that the location domain contains more useful information than the communication domain. $ii$) The performance of NoBal is very close to $LN$, which indicates that the information in the communication domain is submerged by that in the location domain without feature balancing. $iii$) The performance of NoCo is better than CD-SAE, which indicates that the convolutional structure is more suitable to extract features from $\mathcal{A}^L$ and $\mathbf{U}^L$. $iv$) The CD-CNN model achieves the best performance, which indicates that the cross-domain network as well as the co-training algorithm could effectively extract information in unlabeled samples to improve the prediction performance.

\begin{figure}[t!]\centering
	 \subfigure[Calls]{\includegraphics[width=0.8\columnwidth]{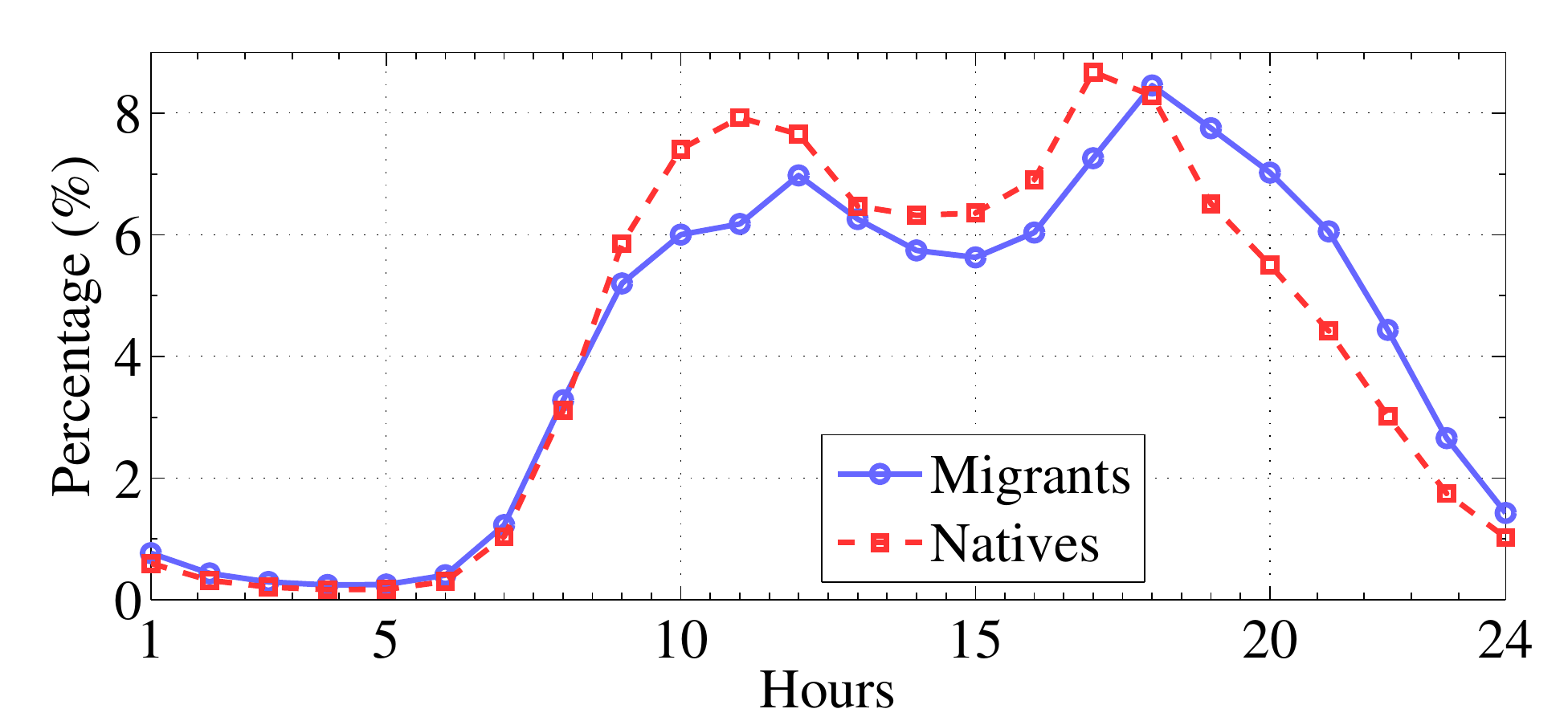}\label{fig:samplePrecision}}\\
\subfigure[SMS]{\includegraphics[width=0.8\columnwidth]{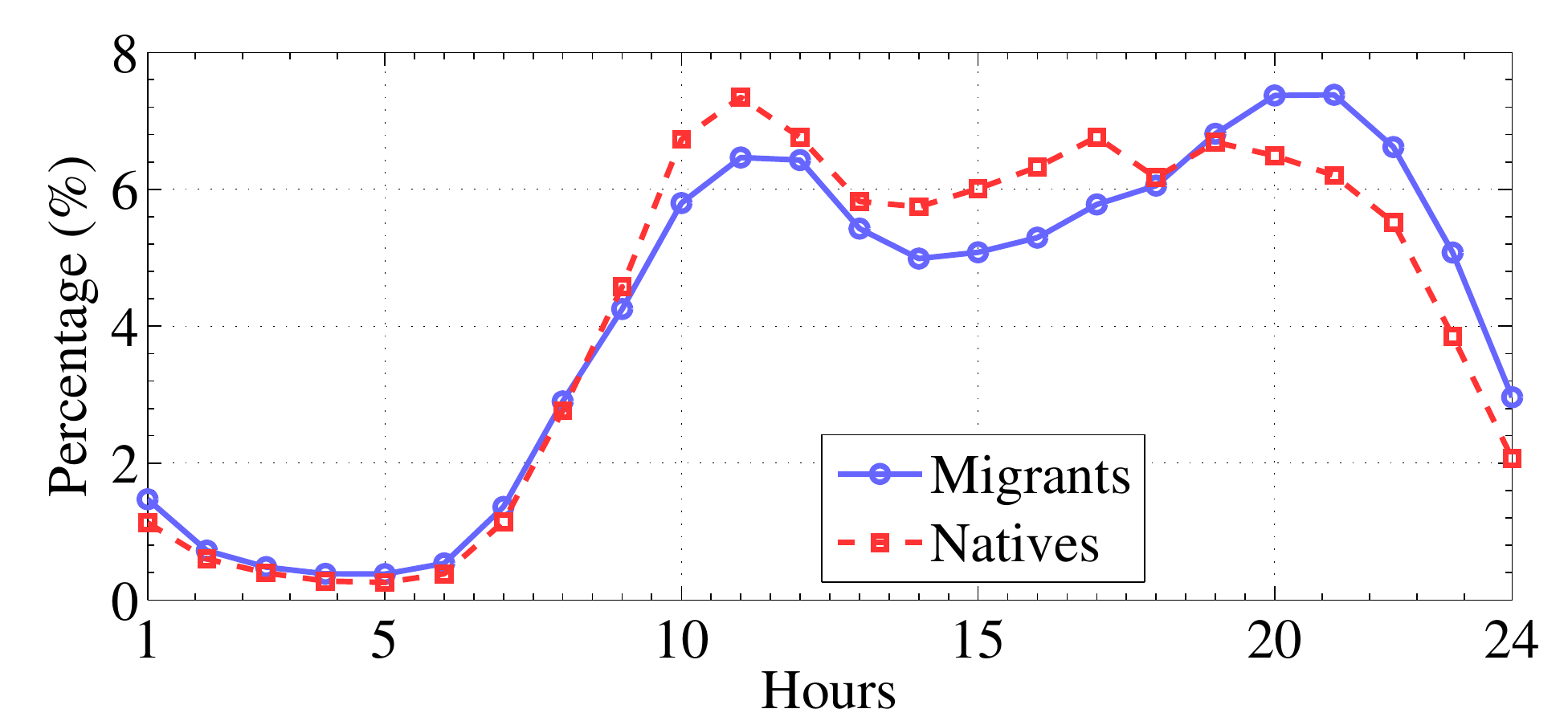}\label{fig:sampleRecall}}
	\vspace{-0.3cm}
	\caption{Temporal distribution of resident communication behaviors.}
	\vspace{-0.3cm}
	\label{fig:behaior_distr}
\end{figure}

It is also worth noting that the classification performance of CD-CNN increases with the collecting days. It may due to that the data set collected from a longer period contains more robust information of residents. Nevertheless, increasing the data collecting time immoderately will incur unaffordable time consumption and money costs. The results indeed point out that the marginal benefit becomes very low when the collecting time is longer than 15 days. So we set the collecting time as 20 workdays per month in practice.

Another factor that impacts prediction performance is the number of labeled samples used in experiments. Figure~\ref{fig:differentSet} gives the results of CD-CNN and the benchmarks on data with varying sizes of labels. As shown in the figure, the performances of the benchmarks, which do not make use of unlabeled samples, sharply degenerate as the labeled samples decrease. On the contrary, the CD-CNN model performs relatively robustly over the small label-size data sets. This is very important for practical applications, where to label samples is often very costly in terms of both time and money.



\section{Applications}
\subsection{Population Census}

\begin{figure*}[t!]\centering
\subfigure[Home Period]{\includegraphics[width=0.75\columnwidth]{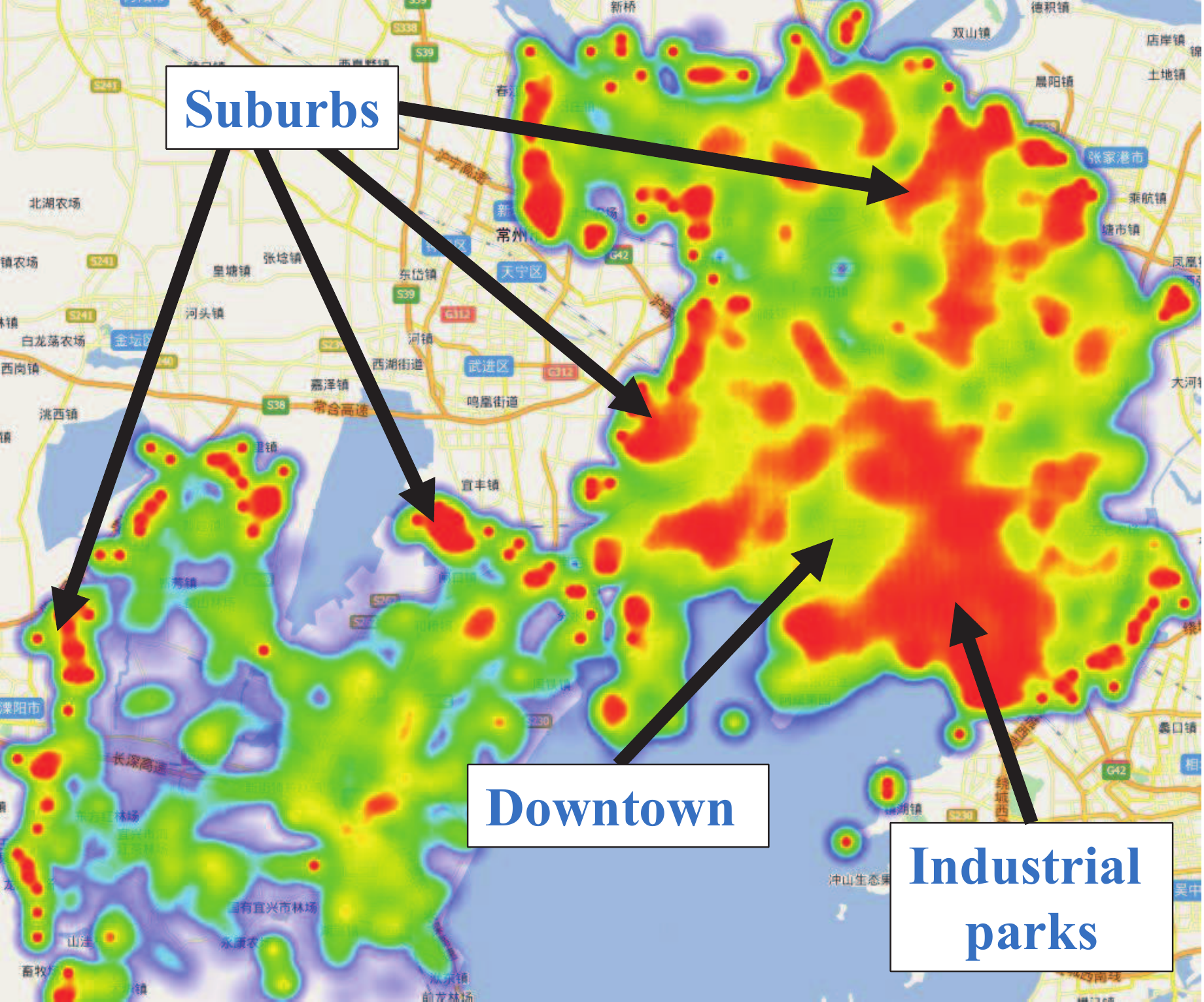}\label{fig:location_night}}
	\hspace{2cm} \subfigure[Working Period]{\includegraphics[width=0.75\columnwidth]{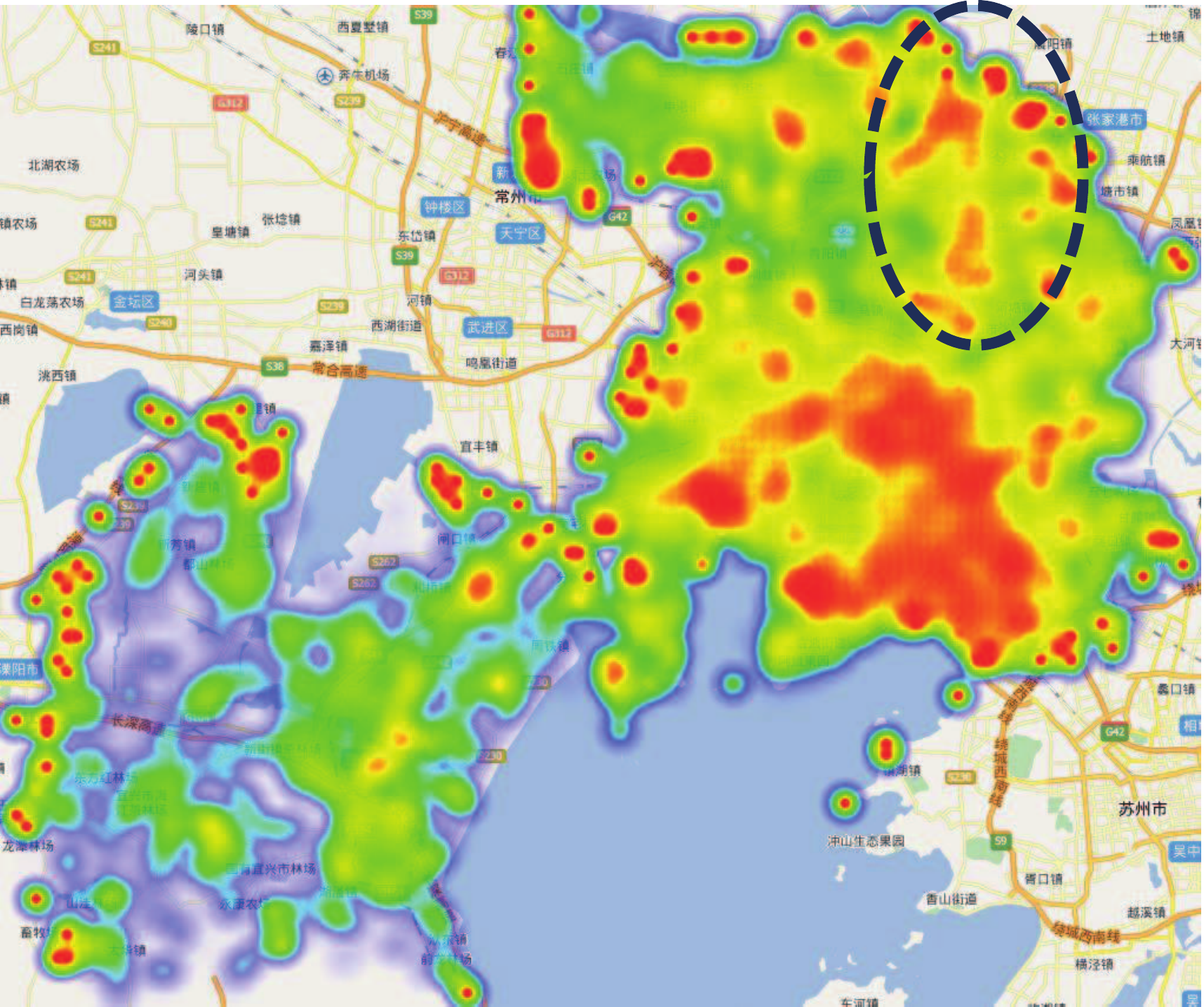}\label{fig:location_day}}
	\vspace{-0.5cm}
	\caption{Spatial distribution maps of migrants in Wuxi.}
	\label{fig:location_distr}
\end{figure*}

The first application of the CD-CNN model is about the fine-grained population census. In the application, we use the mobile phone data of 0.5 million residents collected in 20 workdays as well as 30 thousands questionnaire labels to train a CD-CNN model. We then use the well-trained model to classify all of the five million residents in Wuxi as natives or migrants.

The result shows that 35\% residents in Wuxi are migrants, and the remaining 65\% are natives. We then compare this result with the population census data of Wuxi.  As reported by the Statistical Yearbook of Wuxi~\cite{wuxitongji}, at the end of 2014, the people who lived in Wuxi but had household registrations\footnote{Hukou System: \url{https://en.wikipedia.org/wiki/Hukou_system}} out of Wuxi occupied 28\% of the total population. Furthermore, in the last decade, about 20 thousands migrants translated their household registrations from other places into Wuxi, which occupied 4\% of the total population. Combining the two statistics, the migrants occupy about 28\% + 4\% = 32\% of total population of Wuxi, which is very close to the result inferred by our model. Since our model is able to deliver predictions in a timely and economical way, it could be a valuable complement to traditional infrequent population census.

Moreover, a model-based method could offer a microcosmic and fine-grained behavioral analysis for migrants and natives. Figures~\ref{fig:behaior_distr} and~\ref{fig:location_distr} demonstrate two cases. Figure~\ref{fig:behaior_distr} gives a temporal distribution comparison of call and short message volumes between migrants and natives. As shown in the figure, the evening peak of calls and short messages for migrants is later than the natives. This phenomenon might be due to: $i$) the family of many migrants, such as parents and children, are very likely not in Wuxi, so the migrants need to connect their family through calls and short messages at night; $ii$) migrants of a city usually have greater life pressures than natives, and therefore have to work overtime or handle works during off hours through calls or short messages.

Figure~\ref{fig:location_distr} plots maps of migrant distributions during the home and working periods (the home period is from 19:00 to 7:00 of the next day, and other time is the working period) in Wuxi. The color of the map expresses the proportion of migrants to total residents in an area --- the redder, the higher. As shown in the map, two types of areas have higher migrant proportions: $i$) the areas surrounding the downtown, especially in the industrial parks; $ii$) the suburbs of the city. {These distributions are accordant with intuitive knowledge: First, it is a common phenomenon for cities with a large number of migrants that native residents lived in the downtown areas and migrants lived in the areas surrounding the downtown because most of places in downtown have been already occupied by natives~\cite{outlander}. Second, the well-developed manufacturing industry in Wuxi attracts vast migrants working in the industrial parks of the city. Third, many migrants choose suburbs as residences might due to suburbs have lower housing cost. Finally, compared the two maps, we can found that many migrants, especially lived in the area circled by dashed lines, leave their residences and work in other areas in the working period.}

Based on the results offered by Figs.~\ref{fig:behaior_distr} and~\ref{fig:location_distr}, the urban planning and social welfare departments of the Wuxi government could make proper housing construction plans and social welfare policies to help migrants to improve their living conditions, which is critically important to attracting more talents to migrate to Wuxi.

\begin{table}\centering\small\caption{Leaving rate for different Sigmoid outputs.}\label{tab:home_returning}
\begin{tabular}{c|ccccc}
  \hline
  \toprule
  Conf. & [0, 0.2] & [0.2, 0.4] & [0.4, 0.6] & [0.6, 0.8] & [0.8, 1]\\  \midrule
  Leav. &0.16 & 0.32 & 0.51 & 0.71 & 0.94\\   \bottomrule
  \hline
\end{tabular}
\vspace{-0.3cm}
\end{table}

\subsection{Home Returning Prediction}

The second application is hometown returning prediction of residents during holidays. In major holidays, such as Christmas in Europe and North America as well as Spring Festival in China, migrants are very likely to leave the city they work and return to their hometowns. Therefore, we can use the confidence of classifying a resident as a migrant to predict the hometown returning probability of a resident during a major holiday.

We use the mobile phone data collected during Spring Festival of 2014 (January 31 - February 6, 2014) to verify the hometown returning prediction idea. We label residents as ``leaving'' in the Spring Festival holiday if they have regular mobile phone records in ordinary days but have no record in the holiday. Table~\ref{tab:home_returning} lists the leaving proportions of residents (denoted as ``Leav.'') with different confidences of being classified as a migrant (denoted as ``Conf.''), {\it i.e.}, the Sigmoid output of the LR classifier in the output subnetwork. As shown in the table, the leaving probability of residents is in direct proportion to the confidences. In this way, we could directly use the migrant confidence to predict leaving probability of a resident in a coming Spring Festival holiday. Based on this prediction, we can develop applications to recommend journey/hometown-returning related services to residents, such as ticket booking, express and remittance services, which are very valuable for a business purpose.

\section{Related Works}

Native and migrant studying is a longstanding topic in sociology research areas~\cite{castronova2001immigrants}. Most existing works study the characteristics of natives and migrants, such as cultural~\cite{cultural_capital}, ethnic~\cite{rathelot2014local}, and marriages~\cite{marriages}, based on questionnaires and statistical data. In recent years, social network data are introduced to this area, and research focus of~\cite{yang2017indigenization,windzio2013we} is on analyzing the behaviors of a person with a given native/migrant label. The data-driven methods for identifying whether a person is a migrant or native are yet rarely seen.

In the computer science area, many social computing~\cite{social_computing} works devote to infer profile attributes of people using various data obtained from Internet browsing behaviors~\cite{Internet}, social network behaviors~\cite{tweeting_contains1,tweeting_contains2,tweeting_contains3}, friendship relations~\cite{friendships1,friendships2}, check-in locations~\cite{Inferring}, communication records~\cite{telephone,mobile_phone}, {\it etc}. To the best of our knowledge, this paper is among the earliest studies on identifying native/migrant profile from both online (communications) and offline (locations) human behaviors.

Cross-domain knowledge fusion is a core problem solved by this paper. Related works include multiple modality data fusion~\cite{datafusion} and multi-view learning~\cite{tao}. Many deep learning models are proposed to extract and fuse knowledge from multiple modality data sets~\cite{MDL_icml,MDL_nips}, since deep neural networks have flexible model structures and a powerful representation learning ability. But they did not pay much attention to the dimensionality imbalance problem as we did in this paper. Typical multi-view Learning methods include co-training~\cite{cotraining,co_training_analyzing}, multiple kernel learning~\cite{mkl,mkl2}, subspace learning~\cite{kcca}. For instance, \cite{dtl} uses a co-training algorithm to train a deep neural network for person re-identification. Since the co-training model natively supports semi-supervised learning, we incorporate it into our knowledge fusion framework. In general, the proposed model processes a more complex cross-domain knowledge fusion scenario, where the raw data are high-dimensional, heterogeneous, imbalanced, and with incomplete label information.

\section{Conclusions}
In this paper, a deep learning enabled partially supervised cross-domain knowledge fusion model is proposed to infer the native/migrant attribute of residents from the mobile phone signaling data set with incomplete labels from questionnaires. Specifically, the proposed model uses CNN to extract resident features from both the travel and communication behaviors. The cross-domain knowledge is then fused using a dimensionality balancing mechanism in the network structure as well as a co-training scheme in the network training steps, by which the partially supervised learning is also enabled naturally. The superior performance and value-in-use of the proposed model are demonstrated by experiments over real-world data sets and two interesting applications.

\bibliographystyle{aaai}

\end{document}